# Photonic Fabric Platform for AI Accelerators

Jing Ding
Celestial AI
Santa Clara CA
jding@celestial.ai

Trung Diep
Celestial AI
Santa Clara CA
tdiep@celestial.ai

## ABSTRACT

This paper presents the Photonic Fabric™ and the Photonic Fabric Appliance™ (PFA), a photonic-enabled switch and memory subsystem that delivers low latency, high bandwidth, and low per-bit energy. By integrating high-bandwidth HBM3E memory, an on-module photonic switch, and external DDR5 in a 2.5D electro-optical system-in-package, the PFA offers up to 32 TB of shared memory alongside 115 Tbps of all-to-all digital switching. The Photonic Fabric™ enables distributed AI training and inference to execute parallelism strategies more efficiently.

The Photonic Fabric removes the silicon beachfront constraint that limits the fixed memory-to-compute ratio observed in virtually all current XPU accelerator designs. Replacing a local HBM stack on an XPU with a chiplet that connects to the Photonic Fabric increases its memory capacity and correspondingly its memory bandwidth by offering a flexible path to scaling well beyond the limitations of on-package HBM alone.

We introduce CelestiSim, a lightweight analytical simulator validated on NVIDIA H100 and H200 systems. It is used to evaluate the performance of LLM reference and energy savings on PFA, without any significant change to the GPU core design. With the PFA, the simulation results show that up to 3.66x throughput and 1.40x latency improvements in LLM inference at 405B parameters, up to 7.04x throughput and 1.41x latency improvements at 1T parameters, and 60-90% energy savings in data movement for heavy collective operations in all LLM training scenarios. While these results are shown for NVIDIA GPUs, they can be applied similarly to other AI accelerator designs (XPUs) that share the same fundamental limitation of fixed memory to compute.

## CCS CONCEPTS

• Hardware • Emerging technologies • Analysis and design of emerging devices and systems • Emerging architectures

## KEYWORDS

Interconnect, photonics, memory, machine learning system performance

## 1 Introduction

Over the past decade, rapid advancements in Artificial Intelligence (AI), particularly Generative AI (GenAI), have highlighted significant challenges in current hardware's ability to keep pace with model scaling. While the model sizes have grown exponentially with many open-source and proprietary models now exceeding multiple-trillion parameters, hardware capabilities remain limited by comparatively linear scaling laws. Photonics is widely viewed as a promising technology for the next generation of high-speed, high-bandwidth interconnects, and its comparative benefits over more traditional electronic solutions are forcing substantial changes in production-scale environments.

In this paper, we address the critical overheads introduced by interconnects in data center AI workloads, including large language model (LLM) training and inference as well as deep learning recommendation model (DLRM) inference. We introduce the **Photonic Fabric Appliance™** (PFA), a rack-mountable cluster-scale storage system with up to 32 TB of shared memory capacity at full HBM3E bandwidth, complemented by 115 Tbps of all-to-all digital switching with a radix of 16. This appliance integrates with the Photonic Fabric™ to overcome the silicon beachfront constraint that limits the fixed memory-to-compute ratio in conventional XPU accelerators.

To demonstrate these benefits, we use NVIDIA H100 and H200 GPUs as a case study by developing a parameterized analytical simulator tailored for high-level software-hardware co-design studies, tailored to model LLM training and inference processes. The simulator also incorporates energy and power estimates based on data transfer calculations. With this simulator, we demonstrate PFA's performance improvements across AI workloads. Compared to conventional NVLink with NVSwitch networking, the PFA is projected to achieve up to 3.66x throughput and 1.40x latency improvements in inference for a 405-billion parameter model, and up to 7.04x throughput and 1.41x latency improvements for a 1-trillion parameter model. For heavy collective operations in training using scale-out networking technology such as InfiniBand, the PFA can reduce energy consumption by 60–90%. Additionally, the PFA achieves 22.8x higher performance than GPUs with NVLink for DLRM inference. These results indicate the PFA's potential to overcome memory bottlenecks, scale efficiently, and enable next-generation AI deployments on a broad range of AI accelerators. While these results are shown for NVIDIA GPUs, they can be applied similarly to other AI accelerator designs that share the same fundamental limitation of fixed memory to compute.

## 2 Background

Balancing between compute intensity and sufficient memory performance to feed data to the compute units is key to achieving peak efficiency. In this section, we explore the balancing act that takes place to coordinate between compute and memory demands especially across multiple networked nodes and provide motivations for current state-of-the-art networking strategies.

### 2.1 Current Scale-out and Scale-up Networks

Historically, two primary strategies have been used to handle the growing scale of AI workloads: scale-up architectures, which concentrate resources in more capable single-node systems; and scale-out architectures, which distribute workloads across multiple interconnected nodes. Both approaches now face increasing pressure due to the rapid growth in model size and training corpora.

In *scale-up systems*, one equips individual servers with high performance accelerators (XPUs, TPUs, or custom ASICs) and large pools of high-bandwidth memory (HBM), along with substantial CPU-attached DRAM. This design reduces latency for parameter retrieval and activation storage by placing more memory and compute resources on a single node. However, even sophisticated node-level memory hierarchies—leveraging HBM stacks, large on-die caches, and NUMA-balanced DRAM configurations—have inherent scaling and efficiency limitations. Similar constraints apply to near-memory or in-memory compute, or packaging techniques like 3D stacking and chiplets. Ultimately, the scale of modern AI workloads pushes beyond the feasible capacity, bandwidth, and cost-effectiveness of the most advanced single-node architectures.

For these tasks, a single node, even a heavily provisioned one, just cannot hold all parameters, activations, and cached artifacts. For this, one requires *scale-out architectures*, which partition the model and data across many nodes interconnected through high-performance fabrics, such as InfiniBand and RoCE-based Ethernet. The cost of this increased capacity is the introduction of complex memory access patterns and significant inter-node communication overhead. Models with large attention mechanisms, or large embedding tables, generate bursty, high-bandwidth traffic that can overwhelm network links and switches. Ensuring sustained performance at scale requires careful coordination between memory placement, parallelization strategies, and routing algorithms that mitigate congestion and load imbalance. Techniques such as hierarchical parallelism by combining tensor, data, pipeline, or expert (agent) partitioning become essential. Software-managed caching and compression schemes can reduce per-node memory footprints and network load, while integrated collective libraries co-tuned with the network stack minimize synchronization overheads.

### 2.2 Illusion of a Scale-up Network

As shown in Table 1, scale-out and scale-up networks serve fundamentally different operational goals. Using remote direct memory access (RDMA) semantics (implemented over InfiniBand or Ethernet RoCE) to emulate a scale-up environment on top of a scale-out architecture has limitations. InfiniBand's two-sided verbs (e.g., send and receive) require both communication end points, while RDMA's one-sided verbs (e.g., read and write) require only the source communication point, retrieving or placing data in remote memory without notifying or involving the target. RDMA verbs follow a non-blocking, asynchronous I/O model by issuing a completion signal in the completion queue. The capability to implement RDMA verbs on InfiniBand or Ethernet is made possible by introducing custom software into the RDMA-enabled NIC (RNIC). Much of the networking software that is traditionally executed in a kernel on a CPU can be bypassed by adding data buffers in the RNICs to transfer data without involving the CPUs nor the GPUs to remove network software stack overhead. In a Clos-style, multi-stage, non-blocking networks, overall throughput typically scales with the number of active bandwidth-sensitive traffic flows, yet congestion remains a concern: latency-sensitive traffic is not intrinsically protected from bulk bandwidth-intensive traffic. The use of virtual lane and virtual lane arbitration help to provide per-flow performance differentiation, but multiple factors as well as the imbalance between bandwidth and latency sensitive traffic can pose fairness issues that make achieving low latency and high bandwidth at the same time difficult.

| Characteristics | Scale-out Network | Scale-up Network |
|---|---|---|
| **Scalability** | Many millions | Hundreds |
| **Latency** | Higher | Lower (flit-based) |
| **Bandwidth** | Lower | Higher |
| **Data Delivery** | Best effort | Deterministic |
| **Reliability** | Lossless | Retry possible |
| **Distance** | Long distance | Short point to point |
| **Software** | Ubiquitous | Custom |
| **Cost** | Cheaper | More expensive |

Table 1: Scale-up vs. scale-out networking characteristics.

### 2.3 Memory Demands of AI Workloads

Training LLMs involves frequent collective operations that synchronize parameters and gradients across thousands of XPUs. These operations generate bursty, high-bandwidth traffic patterns that can saturate memory and network resources, making consistent performance and throughput difficult to sustain. The challenges of scaling up LLMs to tens of thousands of XPUs, as demonstrated in large-scale deployments like Meta's RoCE-based backend clusters, underscore the complexity of routing and congestion control. Load imbalances, microburst, and low-entropy traffic patterns all demand careful codesign of hardware and software stacks.

Although inference does not involve gradient updates, it still requires rapid access to large parameter sets, can struggle with latency and bandwidth constraints, and requires adaptable load balancing. Hardware like the NVIDIA H100 GPU provides 989 TFLOPS of dense FP16 compute and a memory bandwidth of 3350 GB/s, implying a peak arithmetic intensity near 295 FLOPS/byte. Balancing compute and memory operations here is complicated by how arithmetic intensity fluctuates with batch size, context length, and the prefill or decoding phases of LLM inference.

Figure 1 characterizes the arithmetic intensity of LLaMA-70B inference in FP16 precision, demonstrating the contrasting behaviors between the prefill and decode phases. In the **prefill phase** (left), arithmetic intensity scales with batch size and initially increases with input length, reflecting higher compute requirement as workloads grow. However, beyond an input length of ~10,000 tokens, arithmetic intensity begins to decline. This is due to the growing dominance of memory-bound operations in the attention mechanism, which cannot be fully mitigated by memory access optimizations. In the **decode phase** (right), arithmetic intensity is significantly lower and exhibits a different trend. It increases with batch size but decreases as the key-value (KV) cache length grows. This is driven by the rising cost of KV memory accesses, which scale with sequence length, and quickly dominate the execution time, exhibiting the memory bandwidth bottlenecks in this phase.

These trends emphasize a key systems challenge: the arithmetic intensity of LLM inference workloads varies significantly across phases and input configurations, making it difficult to align with the fixed operational intensity of XPUs. This variability complicates resource provisioning and motivates the need for adaptive runtime scheduling, memory hierarchy design, and architectural support to maintain high utilization across diverse LLM inference workloads.

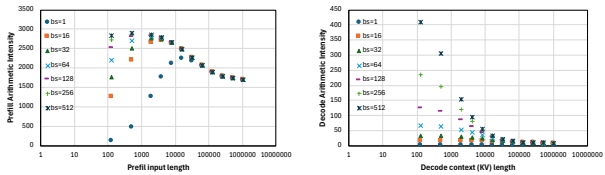

Figure 1: Arithmetic intensity in prefill phase (left graph) and decode phase (right graph).

## 3 Photonic Fabric Appliance™

As both scale-up and scale-out solutions push their practical limits, new interconnect technologies like Celestial AI's Photonic Fabric™ provide a path forward. By providing higher bandwidth densities, more flexible interfacing with compute units, and lower energy per bit transferred, these interconnects can mitigate the complexity of multiple communication collective patterns. On-node photonic integration also provides flexible options for interfacing with larger pools of memory, in turn making it easier to balance performance and avoid capacity-related bottlenecks.

### 3.1 The Need for Photonics Connectivity

The Photonic Fabric Appliance™ (PFA) unlocks a much larger pool of memory and bypasses the memory wall. It is a memory and compute interconnectivity and disaggregation platform operating at 56 Gbps line-rate and providing up to 32 TB of shared memory capacity at full HBM3e bandwidth across a scale-up network of 16 XPUs. The main building block of PFA is the Photonic Fabric Module™ (PFM), shown in 2, which comprises an active photonic interposer codesigned with an advanced Application-Specific Integrated Circuit (ASIC) and two HBM3e stacks and an in-built 8 Tbps network switch in a 2.5D package. Interconnected together, 16 PFMs form a PFA, supporting all-to-all digital switching capability with a radix of 16. The front fiber ports serve as an interface for up to 16 XPUs, as shown in Figure 3, providing all-to-all networking capability along with a unified shared memory address space.

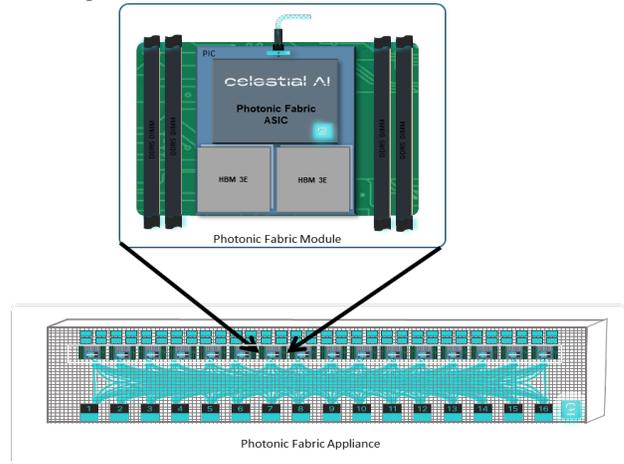

Figure 2: A module in a PFA.

### 3.2 Active Photonic Interposer

The advantage of PFM compared to other solutions relying on photonics for interconnectivity is its exceptional thermal stability, bandwidth density and energy efficiency. The choice of Germanium-Silicon (GeSi) Electro-Absorption Modulators (EAMs) as opposed to Micro-Ring Resonators allows the Photonic Integrated Circuit (PIC) to act as a carrier for a custom ASIC without the link being compromised by thermal dissipation and/or variable temperature gradients. In addition, unlike Mach-Zehnder Modulators, which are also thermally stable, EAMs offer the benefit of compact sub-100μm sizes, thereby providing extremely high package bandwidths. The use of a separate PIC and ASIC that are co-packaged using standard 2.5D/3D assembly techniques, allows the use of advanced 4- and 5nm CMOS nodes for state-of-the-art co-designed Analog/Mixed Signal (AMS) macros and enables the inclusion of the Serializer/De-serializer (SerDes) within the fabric. It also eliminates the need for Digital Signal Processing (DSP) and drastically reduces overall power consumption when compared to Long-Reach (LR) SerDes.

The second generation of our system demonstrator relying on Photonic Fabric™ remains architecturally the same as the first one,

reported in [10] and [11], but comes with several performance improvements obtained from a series of upgrades at all layers of abstraction, including devices, subsystems and control schemes.

## 3.3 PFA Specifications

The Celestial AI PFA is a rack-mountable cluster-scale appliance that supports up to 32 TB of shared memory capacity at full HBM3 bandwidths along with 115 Tbps of all-to-all digital switching capability with a radix of 16. Each PFA incorporates 16 PF modules, as shown in Figure 2. The PF module is a memory fabric & switch ASIC packaged in a 2.5D electro-optical systems-in-package (SIP) with 2x HBM3E stacks of 36 GBs on a photonic IC (PIC) interposer. Each PF module supports up to 2 TB of DDR5 memory capacity with HBM3E acting as write-through cache for the DDR5.

For the purposes of this work, the PFA is configured as an 8 or 16 XPU Cluster, as shown in Figure 3. The PFA has optical port bandwidth of 7.2 Tbps per Photonic Fabric Module, with all-to-all switching totaling 115Tbps per PFA. Additionally, the PFA has embedded memory with each of the 16 attached XPUs having access to the 32 TB of shared DDR5 capacity at HBM3e bandwidths. Additionally, multiple PFAs can be tiered to expand the cluster size to 128 or 256 XPUs with a corresponding increase in the shared memory capacity.

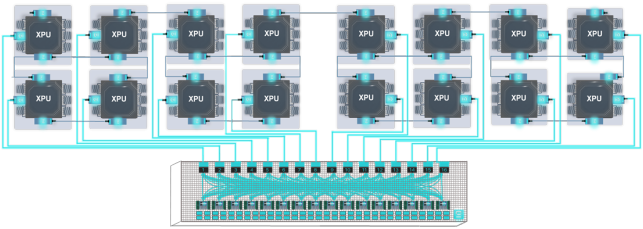

**Figure 3: Connectivity of XPUs with PFA**

## 3.4 Memory Subsystem Architecture

A key benefit of the Photonic Fabric™ is its ability to expand the available memory capacity for each XPU or GPU optically connected to a Photonic Fabric Module™. This overcomes the rigid memory-to-compute ratio observed in most current accelerator designs. For example, while an H100 SXM GPU features a highly capable FP8 tensor core (3,958 TFLOPS) paired with only 80 GB of HBM, its successor, the H200 SXM GPU, increases the HBM capacity to 141 GB due to the limited amount of silicon beachfront available. By contrast, replacing a local HBM stack on an XPU or GPU with a chiplet that connects to a Photonic Fabric Module™ increases its memory capacity to 2 TB without necessarily consuming silicon beachfront. As additional modules are added, each accelerator can seamlessly grow its memory capacity to 4TB or 6TB and correspondingly its memory bandwidth, offering a flexible path to scaling well beyond the limitations of on-package HBM alone.

Another important Photonic Fabric™ implication is the ability to share memory across multiple XPUs or GPUs. As shown in Figure 4, combining 16 Photonic Fabric™ modules along with a crossbar in a Photonic Fabric Appliance™ that can support any connection from any of the Photonic Fabric™ port to any of the Photonic Fabric™ ICs that are directly connected to their respective HBMs, which can provide data at HBM bandwidth while backed by the bigger DDR memory. The ability to share memory across multiple XPUs or GPUs simplifies many of the collective operations required for communicating data that are distributed and stored in individual XPUs or GPUs. For example, in an all-reduce communication collective which aggregates data from multiple XPUs or GPUs and subsequently scatters across the XPUs or GPUs in multiple synchronized steps can be implemented easily by allowing the memory to be locally addressable by each and all XPUs or GPUs that would make up the communication collective.

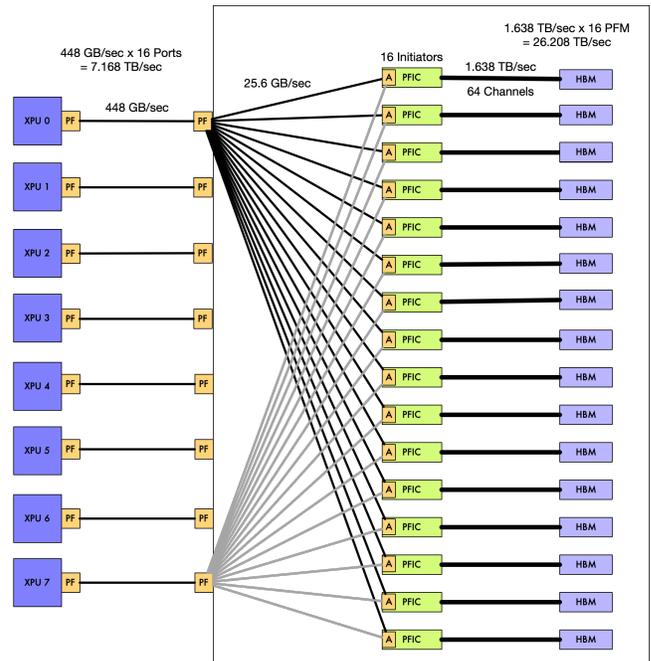

**Figure 4: Memory architecture of the PFA**

## 4 Simulation Framework: CelestiSim

Accurately modeling LLM workloads on emerging architectures — particularly those with disaggregated memory and non-traditional interconnects—requires a simulator that is both hardware-aware and fast enough for iterative co-design exploration. To this end, we developed CelestiSim, a lightweight analytical simulator tailored for transformer-based LLMs running on distributed systems with advanced memory subsystems such as PFA. CelestiSim builds upon foundational analytical modeling techniques [12,14,17] but introduces three key contributions to enable accurate performance and energy prediction for next-generation AI systems:

1. **Support for multi-tier disaggregated memory**: CelestiSim models interactions between on-module HBM3E and photonic-connected DDR5 memory. It incorporates a configurable

**Photonic Fabric Platform for AI Accelerators**

caching strategy to capture the impact of memory hierarchy depth, latency, and bandwidth on LLM execution.

2. **Unified support for both training and inference modeling:** Unlike most simulators that focus on only one mode, CelestiSim models both LLM training and inference. It captures the compute- and memory-bound phases of inference, as well as the interplay of parallelism, communication overlap, and memory access patterns in training—enabling comprehensive analysis of hardware-software interactions across the full LLM lifecycle.

3. **Integrated power and energy modeling:** CelestiSim incorporates analytical energy models to estimate the cost of memory movement, compute operations, and interconnect communication—extending beyond performance prediction to evaluate system-level energy efficiency in LLM operations. This provides insight into the architectural trade-offs of designs like PFA.

CelestiSim enables rapid, yet realistic, exploration of how performance and efficiency scale under novel architectures, including those with photonic switching fabrics. It is validated against empirical measurements on both NVIDIA H100 and H200 GPUs.

## 4.1 Framework Overview

The simulator is a Python-based lightweight analytical performance model designed for high-level co-design of transformer-based LLMs and the hardware systems. It captures interactions among LLM specifications, system configurations, parallelization schemes, and operational modes (training, inference, or power), as shown in Figure 5.

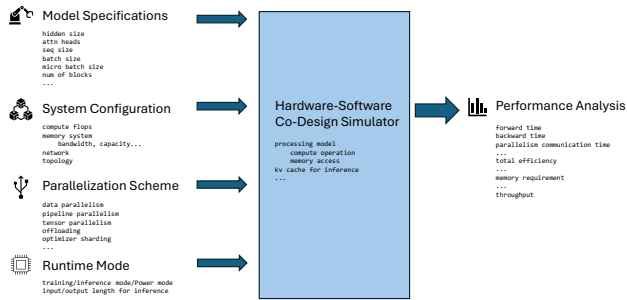

**Figure 5: CelestiSim Framework.**

The simulator adopts the flexible model structure of Megatron [1], compatible with architectures like GPT-2 [18], GPT-3 [19], GPT-4 [20], and LLaMA [21]. Transformer-based models are specified through parameters such as hidden size, attention heads, input sequence length, batch size, micro-batch size, and transformer blocks.

System configurations specify distributed processor systems for matrix operations (e.g., general matrix multiplication) and vector operations. These configurations include FLOPs, input, weight, and output tensor sizes, as well as memory capacities, bandwidths, and efficiencies for different tiers of memory. To match real-world efficiency, we incorporated results from memory access microbenchmark and FLOPs utilization microbenchmark on H100 and H200 GPUs. Since the two GPUs share identical peak compute throughput, we treat their FLOPS utilization as equivalent. These benchmarks reveal fixed latencies for small message sizes and reduced FLOP efficiencies for smaller general matrix multiply (GEMM) operations, as shown in Figure 6. We also observed slightly lower memory bandwidth utilization on H200, likely due to memory controller buffer limitations. These empirical findings are integrated into CelestiSim to support realistic and architecture-aware performance modeling.

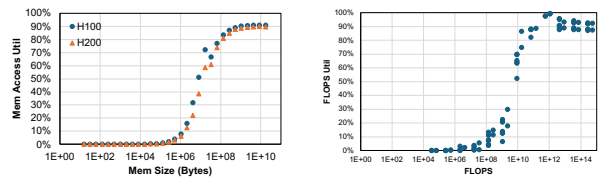

**Figure 6: Memory access bandwidth utilization using different memory transfer sizes (left graph); FLOPS utilization of matrix multiplication on fp16/bf16 (right graph).**

The simulation framework incorporates data parallelism, pipeline parallelism, and tensor parallelism, configured in arbitrary combinations, and incorporates many strategies like data parallelism overlap, 1Forward-1Backward scheduling, sequence parallelism, and decomposing collectives to better hide latency. These optimizations are integrated into synchronous mini-batch stochastic gradient descent with an Adam optimizer. Importantly, CelestiSim factors its analysis out from each layer and ignores scheduling differences between layers. CelestiSim then provides detailed performance estimates and identifies novel hardware-software configurations for efficient LLM execution. It evaluates total efficiency as a function of system and computational efficiencies, enabling comprehensive exploration of design spaces at minimal computational cost.

## 4.2 Power Modeling

The modeling formulates energy consumption as an average over all possible routes in the network, recognizing two main categories of hardware contributors: (1) adapters, such as network interface cards (NICs), GPU or CPU storage adapters, internal PCIe switches, NVLink adapters, and other endpoint interfaces; (2) switches and routers that process, forward, and buffer packets. Each transfer's total per-bit energy is the sum of the source adapter, intermediate switch, and destination adapter costs:

$$E_{total} = E_{s.adapter} + \sum_{i=1}^{N} E_{switch\,i} + E_{d.adapter}$$

where N denotes the number of switches on a particular path. We assume a Clos network architecture comprising multiple racks,

each comprising multiple trays, which in turn comprise multiple GPUs.

Within this framework, there are three principal node-to-node communication scenarios: communication within a single tray (minimal switching), communication within a single rack (inter-tray but intra-rack routing), and inter-rack communication (involving multiple switches, commonly three). There are also two principal offloading communication scenarios: offloading to tray memory (involving GPU, CPU adapters, and potentially internal PCIe or NVLink switches) and offloading to an external data store via a frontend network, typically requiring 4 to 12 switches in the path (e.g., multiple ToR, aggregation, core, and SAN switches). Beyond these five communication scenarios, the path average allows quite general *ad hoc* adjustments for additional routing complexity or network heterogeneity.

We assume that adapters and switch energies are parameterized by estimated per-bit costs: 65 pJ/bit for generic adapters, 35 pJ/bit for generic switches, and 50 pJ/bit for internal NVLink communication [28, 29, 30, 31]. Similarly, we assume energy costs of 5 pJ/bit for photonic transceivers, 25 pJ/bit for photonic switches, and 10 pJ/bit for intra-tray photonic communication.

CelestiSim integrates the above energy modeling into its high-level analytic performance modeling. It simulates training at scale (e.g. tera- to peta-parameter LLMs) across clusters with thousands of GPUs, providing MFU-optimal parallelism strategies (including sizes of all tensor, pipeline, data parallelism clusters) and the bit counts of each data transfer associated with each of the five communication scenarios. By merging this data with the stochastic power model, we estimate cluster layout distributions as well as the resulting network-level probability distributions that enable aggregations of expected energy costs for entire workloads.

## 4.3 Performance Validations

We proceed to verify the accuracy and effectiveness of the CelestiSim particularly for LLM inference in this section. This verification is conducted by comparing the simulator's predictions with empirical data obtained from running the TensorRT-LLM inference engine in a static batch setting on a cluster of eight NVIDIA H100 GPUs or H200 GPUs interconnected using NVLink and NVSwitch in a DGX box.

We validate simulation framework with the LLaMA-3.1 70B model. For H100, we examine tensor parallel (TP) sizes of 4 and 8; for H200, due to its larger memory capacity, we examine TP sizes of 2, 4 and 8. We consider batch sizes of 1, 16, 32, and 64. To evaluate the impact of sequence lengths, we ran two sets of experiments for each model and TP and batch size configuration:

**Variable Input Length:** The input sequence length varies over eight values—1, 32, 64, 128, 256, 512, 1024, and 2048 tokens—while the output sequence length is fixed at 32 tokens.

**Variable Output Length:** The output sequence length varies over seven values—32, 64, 128, 256, 512, 1024, and 2048 tokens—while the input sequence length is fixed at 512 tokens.
In total, we consider 180 configurations for the 70B model.

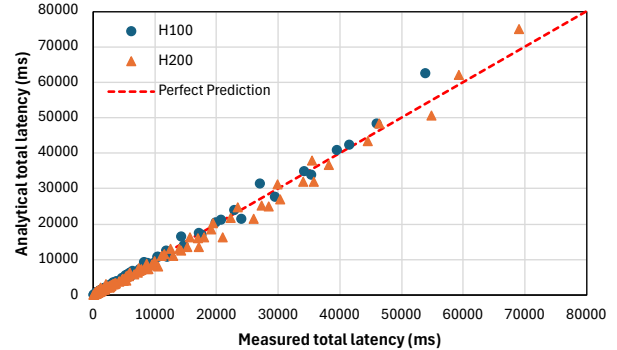

**Figure 7: Validation of CelestiSim predictions using the LLaMA-3.1 70B model.**

For each configuration, we measure the end-to-end execution times of the inference process with both the prefill and decoding phases and use CelestiSim to predict execution times. Figure 7 presents these results. Across all tested batch sizes, input sequence lengths, and output sequence lengths, CelestiSim achieves a mean absolute percentage error (MAPE) of 7.57% and an $R^2$ value of 0.99, indicating strong predictive accuracy and consistency. The simulator also faithfully captures the performance impact of varying TP sizes, including communication delays and synchronization overheads typical in multi-GPU systems. While predictions for the H200 GPU tend to slightly underestimate execution time compared to H100, this trend aligns with the microbenchmark results in Figure 6. Specifically, H200 exhibits marginally lower effective memory bandwidth. These architectural differences are reflected in CelestiSim's performance modeling, validating its ability to capture hardware-level characteristics.

## 5 Power Savings of LLM Pre-training

Power modeling with the CelestiSim demonstrates that migrating from conventional Ethernet-based Clos topologies to the Photonic Fabric[TM] can significantly diminish energy consumption in large-scale pretraining workloads. Across a spectrum of model and cluster configurations, the analyses consistently indicate approximately 60-90% reductions in communication-related power expenditures. These gains appear robust to variations in model size, cluster scale, and the specific blend of parallelization techniques employed.

Bandwidth-intensive tensor parallelism (TP) is critical to achieving high arithmetic utilization for very large models. As model sizes increase, TP communication overheads and associated energy costs grow proportionally. In these scenarios, the Photonic Fabric[TM] curbs this energy usage by up to an order of magnitude, facilitating more cost-effective scaling to the trillion-parameter regime and beyond.

| Model Size | NVIDIA Baseline | PFMM 2TB only | | PFMM 4TB only | | PFMM 6TB only | |
|---|---|---|---|---|---|---|---|
| 1T | 1,026.05 | 190.46 | 18.6% | 190.46 | 18.6% | 190.46 | 18.6% |
| 2T | 1,710.08 | 391.43 | 22.9% | 391.43 | 22.9% | 317.44 | 18.6% |
| 4T | 2,565.12 | 952.32 | 37.1% | 952.32 | 37.1% | 587.15 | 22.9% |
| 7T | 3,665.98 | 1,361.02 | 37.1% | 1,361.02 | 37.1% | 1,361.02 | 37.1% |
| 11T | 6,299.83 | 1,872.90 | 29.7% | 1,872.90 | 29.7% | 1,872.90 | 29.7% |
| 18T | 8,288.55 | 3,038.49 | 36.7% | 2,464.13 | 29.7% | 2,464.13 | 29.7% |
| 26T | 23,381.33 | 4,266.67 | 18.2% | 3,914.32 | 16.7% | 3,174.40 | 13.6% |
| 37T | 29,577.39 | 7,573.27 | 25.6% | 4,951.62 | 16.7% | 4,951.62 | 16.7% |
| 53T | 46,470.41 | 18,678.73 | 40.2% | 13,312.00 | 28.6% | 6,106.34 | 13.1% |
| 72T | 62,655.18 | 25,996.57 | 41.5% | 22,764.70 | 36.3% | 8,112.00 | 12.9% |
| 96T | 75,548.72 | 31,346.30 | 41.5% | 27,449.35 | 36.3% | 9,781.33 | 12.9% |

Tensor Parallelism Power Costs (kJ, batch size=3072)

**Table 2: Tensor Parallelism Energy Costs with Percentages shown relative to NVIDIA baseline**

Memory offloading is another key bottleneck as per-GPU memory remains finite while model parameters and batch sizes increase. Photonic networks reduce the energy cost per bit for offloaded data transfers, enabling more efficient sharding and storage to global memory. One consequence of this is higher training efficiencies for larger models without incurring steep energy penalties that would otherwise be imposed by conventional networking infrastructures. Note that memory offloading costs can drop when a larger model's mean FLOPS utilization benefits from larger tensor parallelism clusters.

Memory Offloading Power Costs (kJ, batch size=3072)

| Model Size | NVIDIA Baseline | PFMM 2TB only | | PFMM 4TB only | | PFMM 6TB only | |
|---|---|---|---|---|---|---|---|
| 1T | 4,032.15 | 1,008.04 | 25.0% | 1,008.04 | 25.0% | 1,008.04 | 25.0% |
| 2T | 8,258.63 | 2,064.66 | 25.0% | 2,064.66 | 25.0% | 2,064.66 | 25.0% |
| 4T | 14,681.48 | 6,996.85 | 47.7% | 6,996.85 | 47.7% | 3,670.37 | 25.0% |
| 7T | 24,696.54 | 10,543.78 | 42.7% | 10,543.78 | 42.7% | 10,543.78 | 42.7% |
| 11T | 97,405.84 | 21,435.26 | 22.0% | 15,357.28 | 15.8% | 15,357.28 | 15.8% |
| 18T | 139,941.82 | 24,864.85 | 17.8% | 21,396.96 | 15.3% | 21,396.96 | 15.3% |
| 26T | 135,843.13 | 33,960.78 | 25.0% | 12,240.31 | 9.0% | 29,280.47 | 21.6% |
| 37T | 280,529.63 | 45,671.06 | 16.3% | 45,671.06 | 16.3% | 17,312.75 | 6.2% |
| 53T | 375,053.85 | 64,211.21 | 17.1% | 47,187.47 | 12.6% | 59,717.19 | 15.9% |
| 72T | 496,516.98 | 82,891.13 | 16.7% | 31,714.04 | 6.4% | 77,414.05 | 15.6% |
| 96T | 901,348.71 | 137,423.38 | 15.2% | 74,409.01 | 8.3% | 99,312.01 | 11.0% |

**Table 3: Memory Offloading Power Costs**

While pipeline parallelism (PP) typically involves fewer total data movement than TP or offloading operations, these communication patterns still benefit. Table 4 shows that the Photonic Fabric still delivers non-trivial efficiency savings, consistently near 80%.

Pipeline Parallelism Power Costs (kJ, batch size=3072)

| Model Size | NVIDIA Baseline | PFMM 2TB only | | PFMM 4TB only | | PFMM 6TB only | |
|---|---|---|---|---|---|---|---|
| 1T | 64.13 | 11.90 | 18.6% | 11.90 | 18.6% | 11.90 | 18.6% |
| 2T | 106.88 | 24.46 | 22.9% | 24.46 | 22.9% | 19.84 | 18.6% |
| 4T | 160.32 | 29.76 | 18.6% | 29.76 | 18.6% | 36.70 | 22.9% |
| 7T | 229.12 | 42.53 | 18.6% | 42.53 | 18.6% | 42.53 | 18.6% |
| 11T | 393.74 | 58.53 | 14.9% | 58.53 | 14.9% | 58.53 | 14.9% |
| 18T | 518.03 | 94.95 | 18.3% | 77.00 | 14.9% | 77.00 | 14.9% |
| 26T | 730.67 | 133.33 | 18.2% | 122.32 | 16.7% | 99.20 | 13.6% |
| 37T | 924.29 | 236.66 | 25.6% | 154.74 | 16.7% | 154.74 | 16.7% |
| 53T | 1,452.20 | 291.86 | 20.1% | 208.00 | 14.3% | 190.82 | 13.1% |
| 72T | 1,957.97 | 406.20 | 20.7% | 355.70 | 18.2% | 253.50 | 12.9% |
| 96T | 2,360.90 | 489.79 | 20.7% | 428.90 | 18.2% | 305.67 | 12.9% |

**Table 4: Pipeline Parallelism Power Costs**

Overall, this power modeling indicates that the Photonic Fabric™ can facilitate balancing performance goals against power constraints in the next generation of large-scale AI training systems.

## 6 Performance Evaluation of LLM Inference with PFA

Inference experiments with the CelestiSim indicate that the PFA can significantly boost LLM inference throughput and reduce latency compared to conventional GPU-based clusters. Across a range of batch sizes and sequence lengths, we see throughput gains of up to 3.66x for LLaMA 405B parameter models and up to 7.04x for the projected 1T parameter models. These benefits hold even as model sizes increase beyond the memory capacity of a single DGX system.

### 6.1 Experiments Setup

We evaluated two hardware configurations (see Table 5):

- **H100-DGX:** A single NVIDIA DGX box with eight H100 GPUs, each offering 80 GB of memory, 1979 TFLOPS at fp8 precision, and 3350 GB/s of HBM bandwidth, and
- **H100 GPUs with Photonic Fabric Appliance™ (PFA):** A novel architecture featuring 32 TB of photonically-accessible memory, and 26800 GB/s of interconnect bandwidth.

| System | Num procs | TFLOPs (fp8) | HBM BW (GB/s) | Network | Memory |
|---|---|---|---|---|---|
| H100-DGX | 8 | 1979 | 3350 | NVLink 900GB/s | 80 GB |
| PFA | 1 | 1979× (1,2,4,8) | 26800 | - | 32 T |

**Table 5. Experimental System Configurations**

For the 405B parameter model, we considered a range of batch sizes and four input-output token lengths, 128 or 4096 tokens each, simulating typical production workloads. On the DGX, we enabled tensor parallelism (cluster size 8) and disabled data and pipeline parallelism (cluster sizes 1). On the PFA, we ran a single configuration with no tensor parallelism.

For the more demanding 1T parameter model, which requires two interconnected DGX-H100 boxes even with fp8 quantization, we enabled both tensor parallelism (cluster size 8) and pipeline parallelism (cluster size 2). We used InfiniBand (100 GB/s bidirectional transfer) for the interconnect, and configured the PFA cluster identically, with both tensor and pipeline parallelism.

### 6.2 Results

Figure 8 presents throughput performance for the LLaMA3.1-405B model, examining four combinations of input-output token lengths across various batch sizes on both DGX-H100 and PFA configurations. Generally, throughput increases with batch size before plateauing as workloads shift from memory-bound to compute-bound conditions. Notably, for DGX-H100, this plateau

<"">Celestial AI</>

primarily results from restricted maximum microbatch sizes due to GPU memory capacity limitations, and the overhead associated with distributed computing. In contrast, the PFA demonstrates substantially higher throughput, benefiting from ample memory capacity and a disaggregated memory pool design that eliminates overhead from tensor parallelism (TP). This advantage is further highlighted in the Model FLOPs Utilization (MFU); for instance, the DGX-H100 reaches only 13.6% MFU at its maximum batch size for the (128, 4096) token scenario, whereas the PFA achieves 49.7% MFU.

Across all tested scenarios, the PFA consistently delivers higher throughput than the DGX-H100. However, throughput gains vary significantly based on the nature of workloads. Specifically, input-output pairs with longer outputs, such as (128, 4096) and (4096, 4096), exhibit larger performance improvements compared to shorter output scenarios like (128, 128) and (4096, 128). This variation occurs because prefill stages are predominantly compute-bound, whereas decode stages are memory-bound. Hence, workloads with shorter outputs, heavily reliant on GPU compute capabilities, show relatively smaller throughput improvements on PFA. Conversely, memory-bound workloads experience significant benefits due to PFA's capability to manage larger batch sizes, eliminate inter-GPU communication overhead, and remove memory access overhead associated with tensor parallelism.

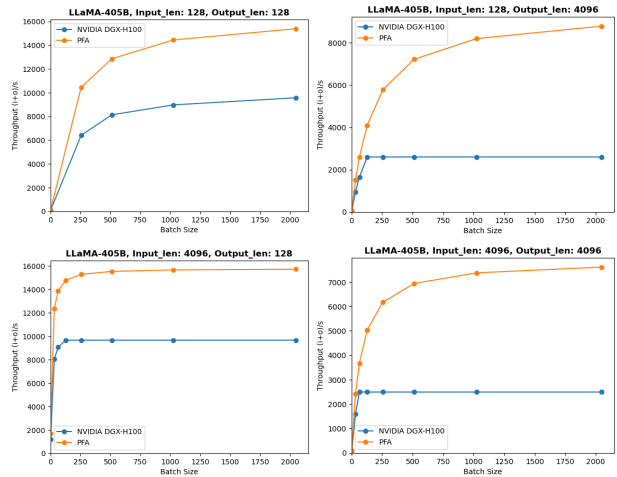

**Figure 8: Throughput results on DGX-H100 and PFA with respect to batch size across different input and output length.**

Figure 9 illustrates throughput and latency speedups provided by the PFA compared to the DGX-H100 across varying compute resource levels, where full compute power is represented by 8 GPUs. We specifically analyze both throughput and latency since both metrics critically influence the overall efficiency and responsiveness of large language model deployments. The PFA achieves notable throughput gains, particularly for memory-bound workloads (e.g., pairs (128, 4096) and (4096, 4096)), demonstrating better throughput even with just a quarter of DGX-H100's compute resources. Regarding latency at batch size 1, the PFA consistently shows improvements at full compute power. When using only one GPU (one-eighth of the total compute power), the input-output pair (4096,128) exhibits limited latency improvement. This is due to the prefill stage significantly dominating the inference time at reduced compute capacity, overshadowing the decoding-time improvements provided by the PFA. On the other hand, even with much smaller compute power, pairs of (4096, 4096) and (128,128) continue to exhibit strong latency improvements because the reductions in decoding times substantially outweigh the marginal increases in prefill durations. In summary, the PFA exhibits throughput improvements of up to 3.66x and latency reductions of up to 1.40x for the 405B parameter model compared to DGX-H100 (Figure 9).

Expanding to larger models, Figure 10 demonstrates even greater benefits, with the PFA achieving throughput improvements of up to 7.04x and latency reductions of up to 1.41x compared to two interconnected DGX-H100 systems for the 1T parameter model. These results underline the scalability and substantial performance advantage of the PFA architecture for large-scale LLM inference workloads.

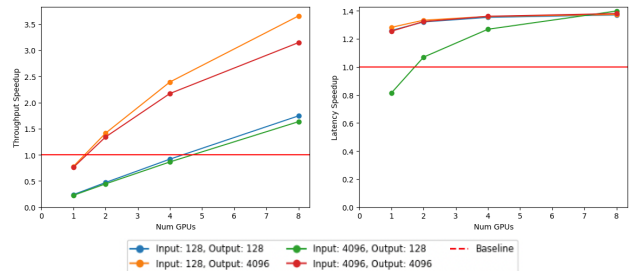

**Figure 9: Left: Throughput speedup and Right: latency speedup, using PFA on LLaMA3.1-405B model over DGX-H100.**

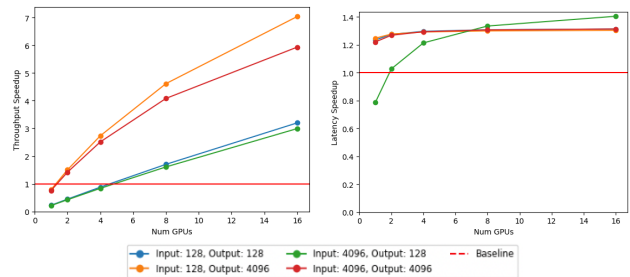

**Figure 10: Left: Throughput speedup and Right: latency speedup, using PFA on 1T model over 2 DGX-H100.**

## 6.3 Discussion

To understand the underlying advantages provided by the Photonic Fabric Appliance (PFA), we analyze the detailed breakdown of latency for key operations during the decoding phase at a batch size of one (Figure 11). The operations listed under "Other" include primarily layer normalization and residual computations. The PFA reduces latency across all operation categories, particularly in communication overhead and layernorm operations. In GPU-based systems like the DGX-H100, the necessity of employing tensor parallelism (TP) contributes significantly to latency overhead. As LLM models scale beyond the memory capacity of individual GPUs, tensor parallelism is essential for partitioning and distributing computations and model parameters. However, this partitioning strategy inherently introduces additional communication and synchronization overhead, negatively affecting overall inference latency.

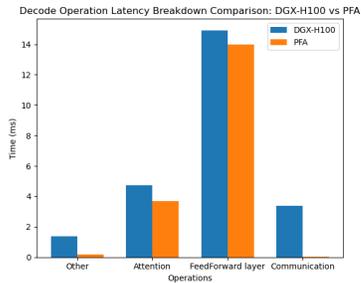

**Figure 11: Operation latency breakdown during decoding phase comparing the DGX-H100 and the PFA.**

Tensor parallelism [1] divides model layers across multiple GPUs, allowing simultaneous computations on separate segments of activations and weights. Subsequently, these GPUs must synchronize and aggregate partial results, often via collective communication operations such as all-reduce. These collective communications introduce fixed latency penalties for small message sizes and bandwidth limitations for large messages, exacerbating overhead as TP scales.

We further illustrate this overhead by profiling the decoding phase of LLM inference under varying TP levels (Figure 12). We use a fixed batch size of 16 tokens, with both input and output sequence lengths set to 128 tokens, evaluating performance across TP sizes of 1, 2, 4, and 8 GPUs. The experiments employed TensorRT-LLM inference engine profiling via NVIDIA Nsight Systems.

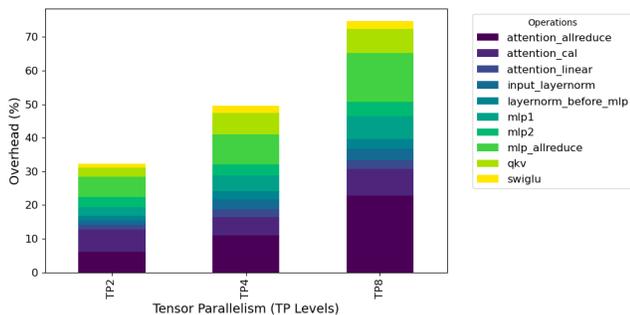

**Figure 12: Breakdown of overhead percentage during decode phase for different TP sizes.**

We calculate "Overhead%" as the fraction of added execution time, relative to a single GPU baseline, normalized by the tensor parallelism (TP) size. Profiling results indicates that overhead% increases as TP size grows. Specifically, all-reduce operations account for 37.68%, 40.10%, 50.02% of total overhead for TP sizes of 2, 4, and 8 respectively. Synchronization further amplifies this penalty, hampering overall performance. Operations like layer normalization, which do not reduce memory access time through partitioning, exhibit elevated overhead.

Besides communication overhead, tensor parallelism inherently induces redundant memory accesses because each GPU must access replicated copies of input/output tensors (illustrated in Figure 13). Such redundancy substantially increases memory access overhead and reduces computational efficiency. Thus, while TP enables deployment of large-scale LLMs, its benefits can diminish due to overhead in communication, redundant memory accesses, and inefficient compute utilization from smaller partitioned workloads.

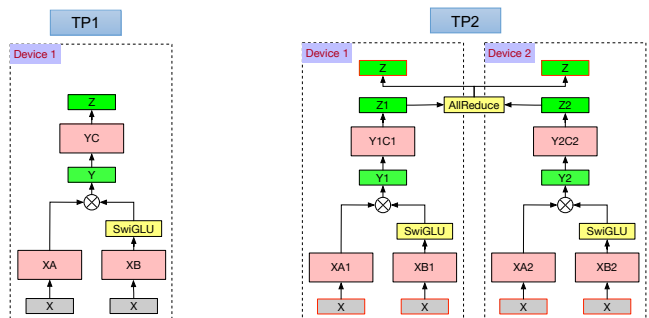

**Figure 13: Illustration of memory access overhead using tensor parallelism for feedforward.**

In summary, the core advantages of the PFA architecture for LLM inference arise from two principal factors:

**Larger Memory Capacity**: Enables efficient utilization of compute resources during memory-bound LLM inference phases, significantly enhancing throughput.

**Reduced Communication and TP Overhead**: By providing a disaggregated memory pool, the PFA drastically lowers communication overhead and redundant memory accesses, thus greatly improving latency-sensitive operations.

In this paper, we specifically highlighted tensor parallelism, as it constitutes a crucial deployment strategy in large-scale LLM inference scenarios and is associated with significant overhead challenges. It is important to note that the advantages of PFA also extend naturally to pipeline parallelism and iterative batching scenarios, particularly those involving memory offloading. Detailed analysis of these additional benefits is deferred to future work.

## 7 Scalability of DLRM Embedding Pooling

For recommendation systems with massive embedding tables, the PFA likewise demonstrates order-of-magnitude performance and efficiency advantages. The Deep Learning Recommendation Models (DLRM) that often drive these systems blend neural networks with often massive embedding tables that can scale into the tens of trillions of parameters [32]. This combination lends itself to exceptionally low arithmetic intensity, as well as complex and often unpredictable communication patterns that underutilize data center infrastructure and lead to subpar performance. The experiments with TorchRec confirm that the PFA can alleviate bottlenecks arising from embedding pooling, one of the main bottlenecks in DLRM inference.

We evaluate against DGX-H100 systems. Our approach partitions the embeddings across multiple GPUs using row-wise parallelism, varying both the number of embedding tables (1, 2, 4, 8, 16, 32, 64) and the batch size (128, 1024, 4096). We fix the embedding dimension at 32 and use pooling factors of 32 or 64. Because each H100 GPU offers 80 GB of HBM, very large embedding tables require distributed storage across tens or even hundreds of GPUs. For example, a 10 TB embedding table occupies 128 GPUs. Under these conditions, the PFA's shared storage—which allows embeddings to be executed entirely in locally addressable memory—and low per-bit photonic energy costs translate into substantial speedups. The simulations indicate an average improvement of 22.8x in comparison to GPUs linked via NVLink, and 28.3x over those connected by PCIe (Figure 14).

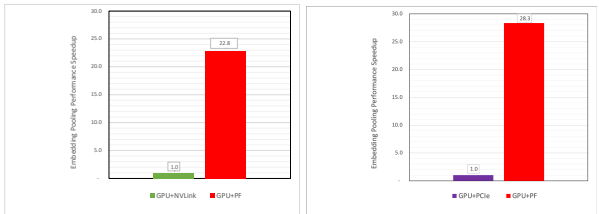

Figure 14: Embedding performance speedup for a 10 TB embedding table

## 8  Related Work

The pragmatic integration of photonics into computing systems has shifted focus toward photonic interconnects. Established semiconductor companies, including NVIDIA [7], have announced interest in on-module silicon photonics for data communication in its high-performance systems, though the company noted the technology is not yet mature for all products. These early-stage efforts validate the growing importance of photonics for memory and interconnect scaling, especially for large language models (LLMs) that demand fast and energy-efficient movement of petabyte-scale key-value data across distributed systems.

However, **evaluating such co-designed architectures**—especially for LLM inference workloads—poses significant modeling challenges. While hardware-oriented simulators can be very accurate, they also tend to be prohibitively slow; system-level GPU simulators such as NVArchSim [8] or network simulators such as SuperSim [9,10] and SST [11] may requires an entire day of computation to model just one second of real-time execution. This makes them impractical for investigating the many parallelization and optimization strategies in modern LLM deployment. Compiler-based models like ParaGraph [9] can be faster but often target only existing hardware and demand extensive engineering to account for the many LLM-specific optimizations. These limitations make them poorly suited for rapid co-design and architectural prototyping—particularly for novel memory systems or inference-heavy scenarios.

To address these challenges, several frameworks have been proposed. Tools like Calculon [12], vTrain [13], and ASTRA-sim [14] provide performance models and simulators to optimize LLM training configurations, focusing on minimizing cost and training time. Other frameworks, such as DeepFlow [17], focus on analytical performance modeling by integrating technology parameters, system architectures, and workload characteristics. DeepFlow utilizes a hierarchical roofline model to predict performance, particularly for matrix multiplication operations, a core component of LLM training workloads. However, these tools often fall short in addressing the dynamic nature of LLM inference. More recent tools incorporate iteration-level simulation and detailed memory modeling to provide better insights into LLM inference performance. LLMServingSim [15], for instance, employs demand paging schemes and computation reuse techniques to achieve feasible simulation times for large-scale inference systems. Vidur [16] combined experimental profiling and predictive modeling to evaluate the end-to-end inference performance for different workloads. Despite significant advancements in tools for LLM performance modeling, accurately capturing the complex interaction between hardware and software in training, inference, and energy costs remains challenging, highlighting the need for precise memory bandwidth utilization models, network simulators for communication overheads, and comprehensive energy analyses.

## 9  Scope of Work and Limitations

While this work demonstrates compelling performance and energy efficiency benefits from integrating the PFA with GPUs for large-scale LLM inference workloads, several limitations and assumptions remain in the current simulation and evaluation framework.

**PFA Hardware Validation and Assumptions**

The evaluation of the PFA relies on predictive modeling using CelestiSim, rather than empirical hardware results. While CelestiSim is validated using microbenchmark data from NVIDIA H100 and H200 GPUs, we apply the same bandwidth utilization and FLOP efficiency models to estimate PFA performance. This assumes comparable compute characteristics and conservative scaling of interconnect bandwidth, rather than modeling novel microarchitectural differences that may exist in the photonic memory interfaces. As physical prototypes of the PFA and its chiplet interface become available, we plan to further validate and refine these assumptions.

**Exclusion of H200 Results in Benchmarking**
Although CelestiSim was validated using both H100 and H200 GPUs, the primary baseline comparison uses H100. This choice reflects a tighter match to the simulation models and greater relevance to current production-scale infrastructure. Including H200 does not change the overall performance trends and may be included in future versions for completeness.

**System-level Modeling Scope**
CelestiSim is designed to capture the dominant hardware-software interactions that influence LLM performance: compute utilization, memory bandwidth saturation, communication latency, and energy costs. However, it does not explicitly model system-level features such as coherence protocols, runtime scheduling semantics, or compiler-specific behaviors. To maintain generality and simulation speed, we make several design-time assumptions:

- **Memory Consistency & Coherency**: We assume that XPUs connected to PFA access memory in a non-coherent manner akin to existing data-parallel or tensor-parallel frameworks (e.g., PyTorch DDP, Megatron-LM). The simulator assumes explicit memory partitioning between compute nodes, with collective communication used to synchronize state.
- **Programming Model**: We follow the Megatron-style programming stack with explicit model parallelism and standard training loop semantics, assuming minimal changes to framework behavior.
- **Scheduling & Workload Coordination**: The simulator decouples per-layer execution and communication, implicitly modeling common scheduling patterns such as 1F1B pipeline parallelism and overlapping mechanisms.

The first generation of the Celestial AI Photonic Fabric products provides a compelling rack-mountable cluster-scale appliance that supports up to 32 TB of shared memory capacity at full HBM3 bandwidths along with 115 Tbps of all-to-all digital switching capability with 16 PF ports. The next generation of the Photonic Fabric products is expected to increase the number of PF ports from 16 to 64 as well as the number of WDM wavelengths from 4 to 8. Using PAM4 signaling, the per link data bandwidth is expected to quadruple from 7.2 Tbps to 28.8 Tbps.

An important implication of the Photonic Fabric is the memory disaggregation that it provides by decoupling the memory from the compute. In addition to the expansion of the memory capacity and the flexible scaling of the memory bandwidth, the Photonic Fabric Appliance mitigates the technology risk from transition of one generation of memory technologies to another. In particular, the support for HBM4 in the next generation of the PFA enables the AI accelerators to continue to achieve higher performance without significant redesign.

## 10 Conclusions

The Photonic Fabric Appliance described in this paper aims to expand the memory and bandwidth resources available to AI accelerators. By integrating high-bandwidth HBM3E memory, an on-module photonic switch, and external DDR5 in a 2.5D electro-optical system-in-package, the PFA offers up to 32 TB of shared memory alongside 115 Tbps of all-to-all digital switching. The simulation results show significant performance speedup and substantially lower energy consumption when the PFA is used in combination with conventional XPUs, particularly for large language models and recommendation workloads. These findings indicate that optical integration and memory disaggregation can help mitigate scaling challenges in AI deployments and can serve as a basis for continued research into more efficient hardware-software co-design for large-scale machine learning.

## ACKNOWLEDGMENTS

We would like to thank the reviewers of this paper. In addition, this work would not have been possible without the many teams at Celestial AI including Photonics group, Packaging group, ASIC design group, ASIC design verification group, Product Marketing group, and the rest of the ML Engineering group. We would like to thank Jonathan Sparling for the work on the LLM pretraining analysis about energy savings.